\documentclass[12pt]{ucl_document}
\usepackage{graphicx,color,psfrag, subfigure}
\usepackage[below]{placeins}
\usepackage{amssymb}
\usepackage{epstopdf}
\usepackage{amsmath}
\usepackage{amsfonts} 
\usepackage{url}
\usepackage{multirow}
 \usepackage[aboveskip=5pt]{caption}

\date{}
\begin{document}
\title[A Theory of Information Matching]{A Theory of Information Matching (TIM)$^*$}
\date{24 May 2012}
\pagestyle{uclpage}
\author[Gorla et al.]{Jagadeesh Gorla, Stephen Robertson, Jun Wang \and Tamas Jambor
 }
\contactdetails{
Jagadeesh Gorla \\
\mbox{\emph{Electronic Mail}: j.gorla@cs.ucl.ac.uk}\\%
University College London\\
Stephen Robertson \\
\mbox{\emph{Electronic Mail}: stephenerobertson@hotmail.co.uk}\\%
Microsoft Research Cambridge\\
Jun Wang\\
\mbox{\emph{Electronic Mail}: jun.wang@cs.ucl.ac.uk}\\%
University College London\\
Tamas Jambor\\
\mbox{\emph{Electronic Mail}:  t.jambor@cs.ucl.ac.uk}\\%
University College London\\
}
\maketitle
\begin{abstract}
In this work, we propose a theory for information matching. It is motivated by the observation that retrieval is about the 
relevance matching between two sets of properties (features), namely, the information need representation and information item representation. 
However, many probabilistic retrieval models rely on fixing one representation and optimizing the other (e.g. fixing the single information need and tuning the document) but not both. Therefore, it is difficult to use the available related information on both the document and the query at the same time in calculating the probability of relevance. In this work, we address the problem by hypothesizing the \emph{relevance} as a logical \emph{relationship} between the two sets of properties; the \emph{relationship} is defined on \emph{two separate} mappings between these properties. By using the hypothesis we develop a unified probabilistic relevance model which is capable of using all the available information. We validate the proposed theory by formulating and developing probabilistic relevance ranking functions for both ad-hoc text retrieval and collaborative filtering. Our derivation in text retrieval illustrates the use of the theory in the situation where no relevance information is available. In collaborative filtering, we show that the resulting recommender model unifies the user and item information into a relevance ranking function without applying any dimensionality reduction techniques or computing explicit similarity between two different users (or items), in contrast to the state\-of\-the\-art recommender models. 
\let\thefootnote\relax\footnote{$^*$ The theory and the mathematical modelling presented in this report has not been published elsewhere. 
However, different applications of the theory are under review.}
\end{abstract}

\section{Introduction}\label{intro}
Information Retrieval (IR) is about \textit{relevance} matching between two sets of properties (features), namely, the information need (query) representation 
and information item (document) representation.  In the past, many authors have attempted to define the various aspects of relevance \cite{mizarro:1} and many 
different models, both non-probabilistic and probabilistic, have been proposed to capture the notion of relevance between them. Some of the influential probabilistic 
models include the classical probabilistic model developed by Robertson and Sp\"{a}rck Jones \cite{robertson:weight}, the Probabilistic Indexing model of 
Maron and Kuhns \cite{Maron}, the language modeling approaches by Ponte and Croft \cite{Ponte98alanguage}, and the risk minimization framework of 
Zhai and Lafferty \cite{zl}. The central problem in all these probabilistic models is the estimation of the probability of relevance, either implicitly or explicitly, 
between a given information item represented by a document and a need represented by a user query.

We note that documents and queries (needs) are typically represented by sets of properties -- we may think of vocabulary terms for example (other examples are discussed below).
 In general, there are two different approaches which the models bring to the formulation of the probability of relevance.  In the first approach, the probability of relevance
 is defined by correlating each document with the information need properties of the users who would judge it relevant, i.e. conditioned by the given document.  This approach 
is called the \emph{document-oriented view} of the probability of relevance \cite{unified1} and includes Maron and Kuhns' Probability Indexing and the language 
models \cite{Maron,Ponte98alanguage}. Whereas in the second approach, the probability of relevance is defined by correlating each user query with the information properties 
of those documents that they would judge relevant, i.e. conditioned by the given information need (query). This approach is called the \emph{query-oriented view} and 
used in the Robertson-Sp\"{a}rck Jones model \cite{robertson:weight}. These two views rely on fixing one variable and optimizing the other, e.g. fixing the information 
need and tuning the document or the other way around, but not both \cite{Microsoft03theunified}. In fact, none of the existing models can use the available relevance 
information on both the document and query in calculating the probability of relevance.

Another important aspect of modern information retrieval modeling is to incorporate properties other than vocabulary terms into the relevance ranking function when computing the probability of relevance. 
For example, previous studies have shown that query independent features, such as PageRank \cite{pagerank}, and the query independent document usage features, such as click-through rates and 
visit frequencies, can be utilized while calculating relevance \cite{Craswell:2005}. As summarized in \cite{Gao:2009, Scholer:2003}, useful information includes query side information such 
as \textit{click-through stream} consisting of all the user queries that have a click on the given document, information from the past and \textit{associated} queries \cite{Scholer:2003}, 
relevant queries for the given document, and information from the set of relevant documents of the query.  However, none of the current probabilistic retrieval models are capable of using all 
the information that might be available.  This may be one of the reasons why learning-to-rank algorithms such as \textit{Lambda Rank} \cite{lambda:2006} perform better than traditional probabilistic 
retrieval models such as BM25 \cite{Robertson:2009}. In large scale web search engines, it is becoming increasingly common to see all available information about the query or/and document being 
used to learn a learning-to-rank model \cite{lambda:2006}, which will then be utilized for ranking the documents based on their relevance to the given query. But in learning-to-rank models 
the results were optimized to general users and personalized ranking is difficult. So, there is no integration of all the information (including user's personal features) in a 
traditional probabilistic retrieval ranking framework, due to lack of a unified theory.

On the other hand, recommendation (collaborative filtering) systems have some similarities to, as well as some significant differences from, information retrieval systems.  In both types of systems, we need to satisfy the requirements of a particular user by offering him/her particular items from a collection.  In the case of information retrieval, we usually start from features (often words), but may also make use of user feedback (relevance feedback).  In the case of recommendation, we usually start from feedback (user ratings) but may also 
make use of features. The most common approach to the task of recommendation relates strongly to information retrieval \cite{wang:tois}.  Given that in many recommendation situations we lack features that could be used directly, it is common to attempt to derive a set of hidden features which might explain the ratings that we observe, and use them to predict new ratings, from either a probabilistic \cite{Hofmann:2004} or non-probabilistic perspective \cite{SVD:Koren}.  These features are usually assumed to describe both users and items, so that both entities may be embedded in the same space -- this parallels the information retrieval situation, where users (in this case user queries) and items both have words as features, and we consider both entities as points in a space defined by words.  The usual assumption in such recommendation systems is that this space is of relatively low dimensionality; although this assumption is by no means universal in information retrieval, it is well represented there in the form of topic models such as PLSI~\cite{Hofmann:1999:PLS:312624.312649} and LDA~\cite{Blei:2003:LDA:944919.944937}.

Thus, in this work, we present a new retrieval theory that can incorporate all the different types of above information into a single model (and as well as personalize the ranking results). 
The basic idea is that the information need and information item are described with their respective properties, potentially from different sets.  The matching for relevance then requires \emph{two separate} mappings between these properties: one from the need to the item properties to identify which item properties are sought by each need, and one in the reverse direction to identify which need properties are `sought' by each item.  The relevance of the information need and information item can be then estimated based on a 
logical relationship of the mappings. The advantage of the unified theory, developed based on this simple idea, is that it is capable of utilizing any available information\footnote{This includes the information about other relevant
documents to the given and document and other relevant queries to the given query.} on both the document and the query in 
determining the probability of relevance. It is, thus, widely applicable to many information retrieval problems that requires the matching between two properties. 
We illustrate its potential and derive two practical algorithms by looking into the ad hoc text retrieval and collaborative filtering problems. On one hand, in text retrieval,
 we show that the theory can handle the situation when there is no relevance information available and derive a practical document ranking function. The TREC evaluation 
shows that the resulting ranking function outperforms some strong baselines. On the other hand, the application of the theory to recommender systems results in a new model 
that computes the probability of relevance between a user-item pair without applying any dimensionality reduction techniques or computing any explicit similarity metric 
between the users or items, in contrast to many state-of-the-art models, e.g. the Matrix Factorization and Dimension Reduction methods \cite{SVD:Koren,Hofmann:2004}, 
the neighborhood-based methods \cite{Shardanand95,Sarwar:2001:ICF:371920.372071}. Our experiments on movie rating data sets demonstrate that it performs significantly 
better than other baselines for the item ranking task.

The remainder of the paper proceeds as follows. In Section 2, we present our unified retrieval theory, and in Section 3, show how to employ the theory to 
derive appropriate ranking functions for both the text retrieval and collaborative filtering tasks.  We then report our experiments in Section 4, and finally 
conclude the paper in Section 5.


\section{Unified Retrieval Theory}\label{gmodel}
\subsection{An Example}
Let us first start with a simple example to demonstrate the idea and insight behind our unified retrieval theory.  We envisage a collection of employers seeking 
candidates to fill their job vacancies, and candidates (job seekers) seeking to find suitable positions. In general, each vacancy is described using its own properties and each 
candidate is described using his or her own properties. There are some properties those can describe only the job vacancy or the candidate but not both, e.g. age, vacancy position 
salary, etc. A candidate with certain properties seeks a job with certain desired properties such as salary, position, etc., and similarly, an employer seeks to fill a vacancy with 
a candidate with certain properties such as qualifications, experience, languages known, etc. A vacancy is filled only if the position has the properties sought by the 
candidate \emph{and} the candidate has the properties sought by the employer for this position.  From a system perspective, to find an ideal match, we have to know the properties 
of candidate and vacancy, and also the properties in the other that are sought by each.

A similar explanation in document retrieval would be an information need with certain properties seeks an information item with certain properties and an information 
item with certain properties seeks to satisfy information needs with certain properties. For example, if a query comes with an identified geolocation, this may (depending 
on the rest of the query) seek a document or page with a nearby geolocation (where the meaning of `nearby' also depends on the rest of the query).  Similarly, a page describing 
a restaurant will probably be `seeking' relatively local people.  On the other hand, we might hypothesize that any query is likely to seek an authoritative 
document (as measured by, say, PageRank).

The basic idea here is that the information need and information item are described with their respective properties, potentially from different 
sets (we could think of these as \emph{vocabularies}, but in principle the vocabulary for need-description is different from that for item-description).  The matching 
for relevance then requires \emph{two separate} mappings between these vocabularies: one from the need properties to the item properties (identifying which item properties are 
sought by each need), and one in the reverse direction (identifying which need properties are `sought' by each item).  
\subsection{A New Hypothesis}
\label{sec:hyp}
Based on this idea presented in the above example, we propose a new hypothesis for IR by making following assumptions: (1) Any information (need/item or document/query) can be 
described by using a set of properties (concepts or features).  (2) The complete set of properties that describe information needs may not be same as those that describe 
information items; (3) An information need seeks an information item with certain description properties and similarly an information item seeks to satisfy an information need with 
certain description properties; (4) All we know about an information need is encapsulated in the properties; therefore we will model the item properties sought by this need as a 
function of the need's properties; and vice-versa.  We will also make the simplifying assumptions: (a) that all properties are binary, and (b) that the two functions 
indicated in (4) are linear and are represented by matrices.

Now, we state an Hypothesis for Information Retrieval as: \\
\emph{``Any information need or information item can be described using a set of properties, called need and item properties respectively.  The relevance between 
an information item-need pair is dependent only on the relationship between the need and item properties that describe them.''}

In order to formulate the hypothesis, let $N$ be the set of $k$ need properties that can describe any information need, where  $N = \{n_1, n_2, \cdots, n_k\}$.  Thus, an information 
need, denoted $I_N$, is described by a vector $\textbf{F}$, of $k$ dimensions, with assumed binary values.
Similarly, let $T$ be the set of $l$ information item properties, where $T = \{t_1, t_2, \cdots, t_l\}$. An information item $I_T$ is represented by an $l$-dimensional 
binary vector $\textbf{E}$. Let $Y$ be an ${N \times T}$ information need seek matrix, representing the information item properties sought by an information need, given 
this need's properties. Each row in $Y$ corresponds to an information need property ($n_f \in N$ where $f \in \{1, \cdots, |N|\}$) and each column is an information 
item property ($t_f \in T$ where $f \in \{1, \cdots, |T|\}$). As a simple case the values of the matrix can be binary, ``1'' if the information need property seeks the 
information item property, ``0'' otherwise. i.e, the values, $n \in N, t \in T, Y[n, t] = 1$ if $n$ seeks $t$, ``0'' otherwise. Similarly, let $Z$ be a 
${T \times N}$ information item seek matrix, representing the information need properties `sought' by an information item, given this item's properties. Each row corresponds to an 
information item property and each columns corresponds to an information need property. The simple binary case, the values, $n \in N, t \in T, Z[t, n] = 1$ if the information item with 
property $t$ seeks to satisfy an information need with property $n$.  Here, $Y, Z$ are property relationship matrices. 

Having defined the two matrices and expressed the relevance hypothesis, we can now put forward another explanation of the matrices.  
Considering $Y$, insofar as it maps needs onto item properties, it implicitly identifies similar needs (which may not start with the same need properties, 
but may be mapped onto the same item properties).  This function of $Y$ would emerge in a relevance feedback environment, from different users identifying the same items as 
relevant to their needs.  Similarly, the matrix $Z$ will identify similar documents, by mapping them onto the same need properties.  These characteristics of the matrices can 
only be expected to emerge in a relevance feedback environment; they will become very clear in the case of collaborative filtering below.  Our ad-hoc retrieval experiments do 
not at this stage include relevance feedback.

\textbf{Relevance under the Hypothesis:} In this paper, we focus on a simple logical model of relevance (on the assumption of perfect knowledge of all 
properties and relationships), while bearing in mind the framework is a general one and other retrieval methods can be derived with different assumptions about the 
relevance. Specifically, the pair $I_N$, $I_T$ is assumed relevant if and only if: (1) all the ``item properties'' sought by the need $I_N$ describe $I_T$; and 
(2) all the ``need properties'' `sought' by the item $I_T$ describe $I_N$.  Under the above hypothesis, we can replace the individual $I_N$ in (1) by its properties, 
and infer the sought item properties by applying $Z$.  Similarly, we can replace the individual $I_T$ in (2) by its properties, and infer the `sought' need properties by 
applying $Y$.  For a simple binary properties case, the relevance conditions can be expressed as follows: (1) $\forall i, j$ if $n_i=1 ~\& ~Y[n_i, t_j] = 1$ then $t_j=1$; 
(2) $\forall i, j$ if $t_j=1$ and $Z[t_j, n_i] = 1$ then $n_i=1$.  
\subsection{Probabilistic Retrieval Model} In order to develop a retrieval model based on the above definition of relevance, we would like to define a complete 
set of need and item properties and determine their values for a given information need or item, and also define the exact relationship matrices $Y, Z$. 
In practice, it is not possible to do so. So, an obvious way to develop a model based on the hypothesis is by defining a restricted set of properties and probabilistically 
modeling their values.  We assume that we have defined $N$, $T$ and derive a probabilistic relevance ranking function to find the probability of relevance 
between $I_N, I_T$ by introducing the uncertainty into the possible $\textbf{F}$, $\textbf{E}$ values for $I_N, I_T$. Similarly, we assume that there is an 
uncertainty over the exact $Y, Z$.  

\subsubsection{Relevance Ranking Function}
The objective of the ranking function is to rank a set of information items for a given information need based on their probability of relevance. 
From the hypothesis we know that the relevance between $I_N, I_T$ can be computed by using $\textbf{E}, \textbf{F}, Y, Z$. So, in order to rank the 
items for a given need, we compute the probability of relevance between any $I_N$, $I_T$, as follows, 
\begin{equation}\label{eq1}
P(R=1| I_N, I_T) = \\ \sum_{\alpha} \sum_{\beta} \sum_{\gamma} \sum_{\delta} P(R=1, \textbf{E},\textbf{F},Y, Z | I_N, I_T) 
\end{equation}
where $R=1$ means relevant, and $\alpha, \beta, \gamma, \delta$ are all the possible binary vectors and matrices of  \textbf{E}, \textbf{F} and $Y$, $Z$ respectively.
From the hypothesis, $\textbf{E}$, $\textbf{F}$, $Y$, $Z$  are sufficient to determine the relevance between the $I_N$ and $I_T$. 
And also, $\textbf{E}$ is dependent only on $I_T$, $\textbf{F}$ is dependent only on $I_N$, and $Y$ and $Z$ are independent of 
$I_N$, $I_T$. By applying Bayesian transformations and independence assumptions, we get 
\begin{equation}\label{eq2}
P(R=1| I_N, I_T) = P(R=1) \\ \sum_{\alpha} \sum_{\beta} \sum_{\gamma} \sum_{\delta} \frac{P(\textbf{E}, \textbf{F},Y, Z| R=1)}{ P(\textbf{E}) P(\textbf{F}) } 
  P(\textbf{E}|I_T) P(\textbf{F}| I_N) 
\end{equation} Here, we assume that the property description value of an item property to $I_T$ is independent of other properties and similarly, 
need property to an $I_N$ need is independent of other properties. We make another assumption that each entry value in $Y$ is independent of other values in $Y$, similarly, 
the entries in $Z$.\footnote{A `need property' seeking an `item property' is independent of other properties and vice-versa.} 
Based on the these assumptions, we can write Eq.\eqref{eq2} as 
\begin{multline}\label{eq3}
P(R=1| I_N, I_T) = P(R=1)\sum_{\alpha} \sum_{\beta} \sum_{\gamma} \sum_{\delta}  \prod_l \prod_m P(E_l, F_m,Y_{lm}, Z_{ml}| R=1)\\ \underbrace{\frac{P(E_l|I_T)}{ P(E_l)}}_\text{Single property score}
   \frac{P(F_m| I_N)}{P(F_m)}
\end{multline}
where $l \in \{1, 2, \cdots, |T|\}$ and $m \in \{1, 2, \cdots, |N|\}$. Eq.\eqref{eq3} is the final probabilistic unified relevance ranking function. 

To explain the behavior of the ranking function above, let us consider that there are two properties in set $T$, where $T = \{I_{t_1}, A\}$, $I_{t_1}$ is a property 
associated with a vocabulary term and $A$ is an ``authority'' property describes whether the information item is authoritative or not. For example, one might use PageRank to 
indicate authority, with a threshold to define a binary property. Similarly, consider one property in $N$, where $N = \{N_{t_1}\}$, $N_{t_1}$ is a term description property.  Now, let 
us assume that the matrices $Y$ and $Z$ as follows, $Y = \begin{bmatrix} 1 & 1 \end{bmatrix}$ and $Z = \begin{bmatrix} 1 & - \end{bmatrix}^{T}$.  $Y[N_{t1}, A] = 1$ means that any 
information need with property $N_{t1}$ seeks an information item with an authority property $A$.  ``$-$'' indicates that the value could be ``0'' or ``1'', meaning that we assume that its value 
does not affect the relevance with respect to this property relationship.  Note that the ranking function in Eq.\eqref{eq3} can use any information about the document (information item) or query (need) by modeling them as properties, defining their relationships through $Y, Z$ matrices and estimating their value for the given information need/item. Thus all information about the individual item, individual need, and other relevant need-item pairs that share property values, is included in determining the relevance, which is an essential for a unified model \cite{unified1}.

\begin{figure}[t]
  \centering
  \subfigure[Description probabilities]{\label{fig:ppr}\includegraphics[width=0.5\textwidth]{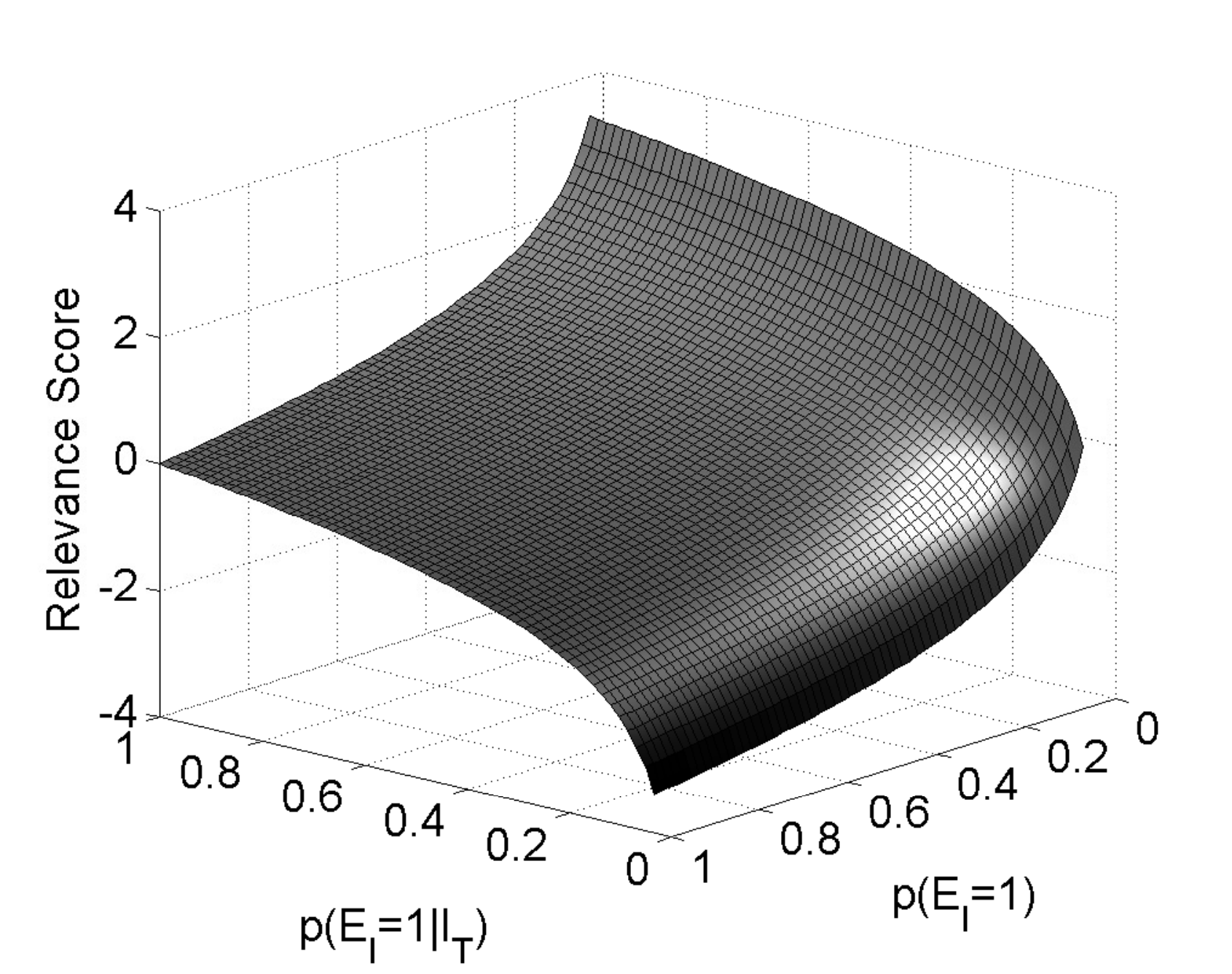}}\\
  \subfigure[Two properties]{\label{fig:tpr1}\includegraphics[width=0.5\textwidth]{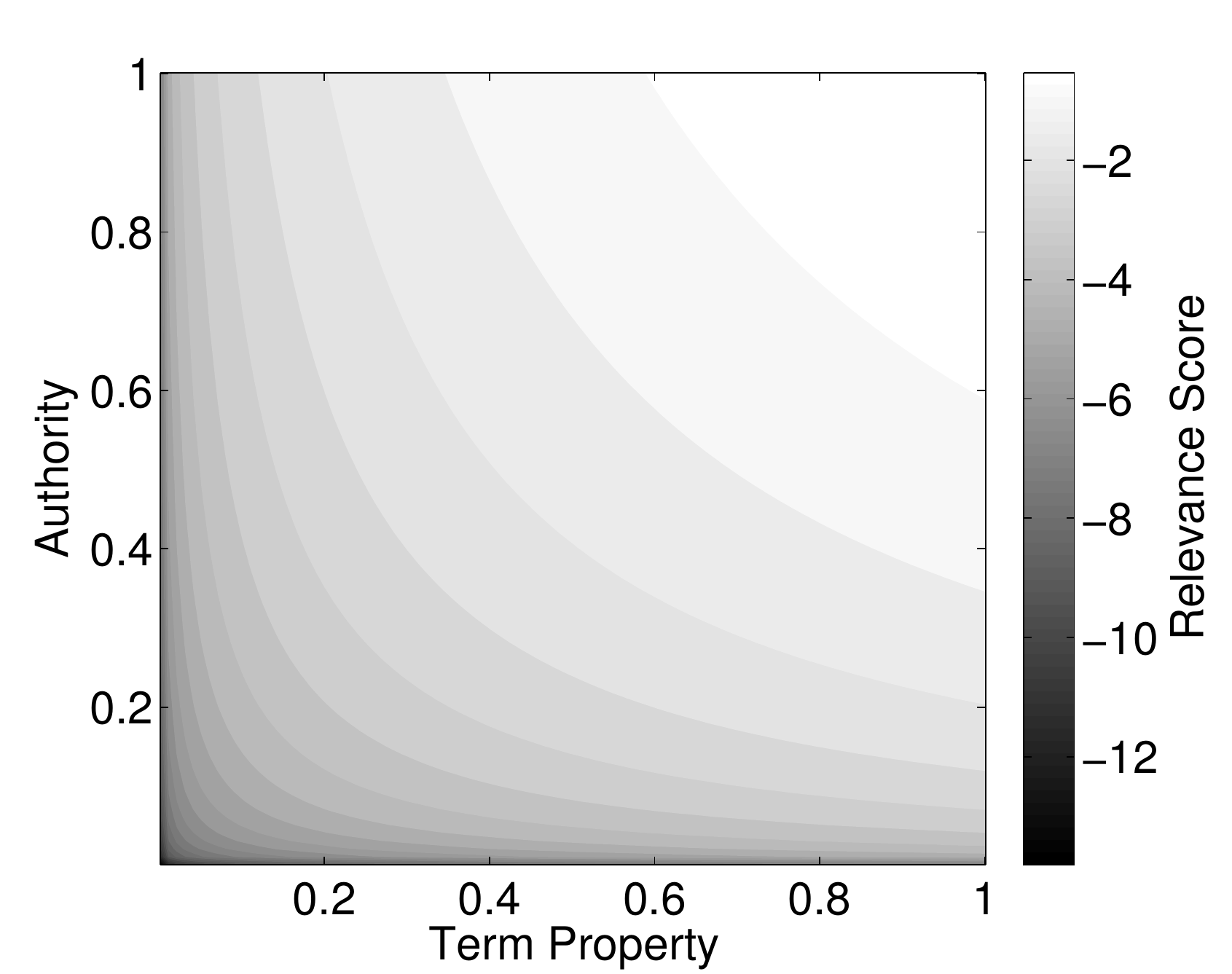}} 
  \caption{The behavior of the Unified Ranking function.}
  \label{fig:plots}
\end{figure}

Now, if we substitute these values of $Y, Z$ into Eq. \eqref{eq3} and assume that the only available information about the information items is $I_T$, then the ranking 
function score depends on ``Single property score'' of both $I_{t_1}$ and $A$ in Eq.\eqref{eq3}. To see how these scores affect the rankings score, we show graph \ref{fig:ppr}.  
The two base axes are the numerator of the property (the probability that the property describes the information items), and denominator (the probability that the 
property describes any information item in general).  The vertical axis is the relevance score as logarithm of ``Single property score'' of Eq. \eqref{eq3}. The maximum 
relevance score is achieved when $P(E_l=1)$ is minimal and $P(E_l=1|I_T)$ is maximal.  This then implies that the property describes very few information items (low $P(E_l=1)$) but 
well describes the particular information item $I_T$ ($P(E_l=1|I_T=1)$ is maximal).  This is what one would expect of a reasonable ranking function.  Graph \ref{fig:tpr1} shows how the relevance score changes when there are two properties (authority and a term property) where $X, Y$ axises describe each ``Single property score'' 
and  $Z$ axis is the sum of logarithmic scores of two properties.  So, the overall relevance score depends on how important the properties are in describing all the information 
items and how well each of them describes the information item (or need). Note that, the relevance component in Eq. \eqref{eq3} includes adding the relevance information on both the 
information item and need which is essential to a unified model. We will use this component in the following applications.

\pagebreak
\section{Applications}
\subsection{Ad-Hoc Text Retrieval}
To develop a text retrieval ranking function using the above unified theory, first, we need to define the property sets $N$ and $T$. 
$N$ (query or need properties) could be a set of properties associated with vocabulary words, query specific properties such as geolocation, query length, etc.
Similarly, $T$ could be a set of vocabulary term properties, document specific properties such as PageRank, url depth, etc.

In a traditional ad-hoc retrieval task, the only available information to the retrieval models is vocabulary terms and their statistics in queries and documents. 
So, to derive a simple ad-hoc retrieval ranking function, we define both $N$ and $T$ as a set of ``$k$'' properties each corresponding to a single vocabulary term. 
We call this set of properties as ``term-description'' properties and represent with $\xi$.  Now, to define the matrices $Y$ and $Z$, 
we define a relationship between the information need and item properties as follows: As $N=T$, an information need with a description property, 
$t$, seeks an information need with same $t$ and vice-versa ($t \in \xi$). Based on the assumption, $Y, Z$ matrices are defined as follows, $Y=Z= M_{|\xi|\times |\xi|}$, 
where $M_{ij} =1$ if $i=j$, ``0'' otherwise and $i, j \in \{1, \cdots, |\xi|\}$.

Following the above relation, the definition of relevance between an information item (document), $d$ and an information need (query) $q$ under the hypothesis reduces to a 
simple relationship where $d$ and $q$ are relevant if and only if $\textbf{E} =\textbf{F}$, i.e. the property description value of all the properties of $d$ must be 
same as the that of $q$. We refer to this relevance relationship as relevance under ``Strict identity'' relation. The reason for this reduction is that we do not need $Y$ and $Z$ for the 
computation of relevance as we know that the same properties should describe both document and query if they are relevant, i.e the description property values for each 
property in $\xi$ must be same for both $d, q$.\\
\emph{\textbf{Ranking Function}}: Now, as per the above definition of \textit{relevance}, the probability of relevance between $d, q$,  $P(R=1 | d, q)$,  can be computed as
\begin{equation} \label{eq4}
P(R=1 | d, q)  \propto_{R}  \sum_{\alpha} \sum_{\beta} \prod_{i} \frac{ P(E_{i}, F_{i}| R=1) } { P(E_{i}) P(F_{i})}  P(E_{i}| d) P(F_{i}| q)
\end{equation}
where $\propto_{R}$ is rank equivalence (constant $P(R=1)$ is ignored) and $i \in \{1, 2 \cdots,  |\xi|\}$.  Eq.~\eqref{eq4} is 
a unified relevance ranking function for ad-hoc retrieval when the same set of properties can describe a document or query. 
Eq.~\eqref{eq4}  uses the information about the description value of each property for the given document and query ($P(E_i| d)$, $P(F_i|q)$), and its value 
in the collection ($P(E_i)$, $P(F_i)$) and the joint probability of property values those describe relevant document-query pairs.  If there is a new relevant pair, 
its information will be added in computing the relevance.  
                 
In traditional TREC collections there is a very little text on query side, so, to implement the ranking function we avoid the estimation of the property values for the given query by making 
the following assumption. 
\emph{Query property description assumption:} As we have very little information (only two to three query terms) to infer the query description property values and the 
fact that each query term is very important in finding the relevant documents, we assume that each property description value corresponding to 
query terms is ``1'' for the given query and others properties are ``0'', i.e. we know the binary vector \textbf{F}. 
Basically, this assumption is similar to an implicit assumption that the query terms are elite to the query and other terms are non-elite as in \cite{RobertsonW:1994}.  
In what follows, we use the terms \emph{elite} and \emph{non-elite} as synonymous with `has the property' and `does not have the property' respectively, 
for either users or items.

Now, let $\mathit{fq}_{i}$ denotes the property value of $F_{i}$ and $F_{i} = \mathit{fq}_{i}$ where $\mathit{fq}_{i}=1$ if $F_{i}$ is elite for $q$ otherwise 
$\mathit{fq}_{i}=0$.
Based on our assumption, Eq.~\eqref{eq4} can be written as 
\begin{equation}\label{eq5}
P(R=1 | d, q) \propto_{R} \sum_{\alpha} \prod_{i=1}^{k} \Big( \underbrace{\frac{ P(E_{i}, F_{i}=\mathit{fq}_{i}| R=1) } { P(E_{i}) P(F_{i}=\mathit{fq}_{i})}}_\text{part 1}
\\ P(E_{i}| d) P(F_{i}=\mathit{fq}_{i}| q)\Big)
\end{equation} 
By applying Bayes' rule to the part 1 to Eq.\eqref{eq5} and  factorizing, we get 
\begin{multline}\label{eq6}
 P(R=1 | d, q) \propto_{R} \sum_{\alpha}  \prod_{\forall i : F_{i} = 0} \frac{ P(E_{i}| F_{i}=0, R=1)}{P(E_{i})P(F_{i}=0)}   
P(F_{i}=0| R=1) P(E_{i}| d) P(F_{i}=0| q) \\ \prod_{\forall i:F_{i}=1}
  \frac{ P(E_{i}| F_{i}=1, R=1) P(F_{i}=1| R=1) P(E_{i}| d)P(F_{i}=1| q) } { P(E_{i}) P(F_{i}=1)}
\end{multline} 
From the \emph{Query property description assumption}, we know the value of each element in \textbf{F}. 
So, we have $P(F_{i}=1| q) =1$ if the term associated with the $i^{th}$ property in $\xi$ is present in  query $q$ and 
$P(F_{i}=0| q) =1$, otherwise. By substituting these values, Eq.~\eqref{eq6} becomes 
\begin{multline}\label{eq7}
 P(R=1 | d, q)  \propto_{R}  \sum_{\alpha}  \prod_{\forall i:F_{i} = 0} \Big( \frac{ P(E_{i}| F_{i}=0, R=1) P(F_{i}=0| R=1)} { P(E_{i}) P(F_{i}=0)}  P(E_{i}| d)\Big) \\ \prod_{\forall i :F_{i} = 1} \Big( \frac{ P(E_{i}| F_{i}=1, R=1) P(F_{i}=1| R=1)} { P(E_{i}) P(F_{i}=1)}  P(E_{i}| d)\Big)
\end{multline}
As defined, a document and a query is relevant if and only if \textbf{F} $=$ \textbf{E}. From the definition, if we know that the property description of a property is ``1'' for the given 
query (i.e. $F_{i}=1$), then the probability that of same property value is ``1'' ($E_{i}=1$) in the relevant set of documents is ``1'',  i.e. $P(E_{i} =1| R=1, F_{i}=1) = 1$, as they have 
the same value in the relevant set. Equally, it is the same for the property value equal to ``0'' where $P(E_{i} =0| R=1, F_{i}=0) = 1$.
 Note that from this assumption, the score of any vector in $\textbf{E}$ of Eq.~\eqref{eq7} is zero if $E_{i}=0$ and $F_{i} = 1$ (or) $E_{i}=1$ and $F_{i} = 0$ for 
at least one $i$. By substituting these values in Eq.~\eqref{eq7}, we get 
\begin{multline} \label{eq8}
 P(R=1 | d, q)   \propto_{R} \prod_{\forall i=1:F_{i} = 0} \frac{ P(F_{i}=0| R=1)} { P(E_{i}=0) P(F_{i}=0)}  P(E_{i}=0| d)
 \\ \prod_{\forall i:F_{i}} \frac{ P(F_{i}=1| R=1)} { P(E_{i}=1) P(F_{i}=1)}  P(E_{i}=1| d)
\end{multline}
where $P(F_{i}=0| R=1)$, $ P(F_{i}=1)$, $P(F_{i}=0)$ and $P(F_{i}=1| R=1)$ in Eq.~\eqref{eq8} can be removed as these terms do not affect the ranking order. We thus get 
\begin{equation}\label{eq9}
P(R=1 | d, q)  \propto_{R} \prod_{\forall i:F_{i} = 0}  \frac{P(E_{i}=0| d)} {P(E_{i}=0) } \\ \prod_{\forall i:F_{i} = 1} \frac{ P(E_{i}=1| d) } {P(E_{i}=1)}
\end{equation}
Eq.~\eqref{eq9} is a ranking function under the Strict identity relation with the \emph{Query property description assumption}.
We ignore the terms with  properties values ``0''  in Eq.~\eqref{eq9} by assuming that the absence of terms represents unknown properties. 
By applying a logarithm transform to the ranking function results in the following 
ranking function:
 \begin{equation}\label{eq10} P(R=1 | d, q) \propto_R \sum_{\forall i:F_{i} = 1} \log \frac{ P(E_{i}=1| d) } {P(E_{i}=1)}
 \end{equation}
The above simplification is similar to ignoring the terms that are not present in query in \cite{robertson:weight,Ponte98alanguage,zl}. 
Note that each property score in ranking function in Eq.\eqref{eq10} has same behavior as shown in the graph \ref{fig:ppr} which is a desired 
characteristic of a relevance ranking function. 

One of the interesting by-products of our model is that the above formula in Eq.~\eqref{eq10} provides a yet another theoretical 
justification of IDF (inverse document frequency) as scoring function \cite{Jones72astatistical}. To see this, let us assume that the property description value of a property 
is ``1'' to a document if the term is present in the document and ``0'' otherwise.  Then, the probability of the property value being ``1'' 
in the collection is, $P(E_i=1) \equiv \frac{n_{ci}}{N},$
where $n_{ci}$ is the number of documents in the collection with the term associated with the $i^{th}$ term-description property and $N$ is total number of documents in the collection. 
From the above assumption, $P(E_i=1|d) \equiv1$ if $F_i=1$ (term $i$ is in the query).  By substituting them in the ranking function in Eq.~\eqref{eq10}, we get
 \begin{equation}\label{idf}
 P(R=1 | d, q)  \propto_R \sum_{\forall i:F_{i}} \log \frac{1}{p_i} =  \sum_{\forall i:F_{i} = 1} \log  \frac{N}{n_{ci}} 
\end{equation} 
Now, the ranking function in Eq.~\eqref{idf} is simply a function of IDF values of the query terms. Essentially, it implies that the IDF score function relies on 
the assumption that a term is elite if it occurs in the document. This is different from the explanation provided by the Robertson-Sp\"{a}rck Jones model, 
where an explicit assumption that the whole collection is a non-relevant set is needed \cite{robertson:weight}.

To implement and test the ranking function in Eq.~\eqref{eq10}, we need to estimate the probabilities  $P(E_i=1| d)$ and $ P(E_i=1)$ for each $i^{th}$ property in $\xi$.
In order to estimate the probabilities, we assume the following generative process where, an author (or a user) will carry out the following process to express their information:
(1) First, a user or author will choose a set of elite properties such that these properties can describe every aspect of the information that they want to express.
(2) Once the properties are chosen, an observable information item or need, is generated by a stochastic function of chosen properties. The uncertainty about the description of 
the property for the information item is injected during this generation process. Now, we know that a  document is generated from a set of 
term-properties. So, the occurrence of a term in a document has a stochastic element associated with the description of its corresponding term-description property. 
Therefore, we compute the probability of $i^{th}$ term-description property value being ``1'' for the given document $d$ as  $P(E_{i}=1| d) \equiv P(E_{i}=1| \mathit{tf}_i)$
 where $\mathit{tf}_i$ denotes the term frequency associated with $i^{th}$ term-description property in document $d$. As we assume that the description of a property for a document is binary, 
from the hypothesis, a property description is ``1'' for some documents in the collection and ``0'' for others. And, $\mathit{tf}_i$ follows one distribution in a set of documents that 
were described by the property and another distribution in second set of documents that were not described by the property. Therefore, we can draw a probabilistic inference about 
the description of a term-description property from its associated term's frequency in the document.

By applying Bayes' rule to $P(E_{i}=1| \mathit{tf}_i)$, we get 
\begin{equation} \label{eqest}
 P(E_{i}=1| \mathit{tf}_i) =  \frac{P(\mathit{tf}_i|E_{i}=1) P(E_{i}=1)}{\sum_{E_{i}\in\{0,1\}}P(\mathit{tf}_i|E_{i}) P(E_{i})}
\end{equation}
For simplicity, we use query terms to represent properties that describe the query ($F_{i} = 1$ when $q_{i} = 1$).  
By substituting Eq.~\eqref{eqest} back in Eq.~\eqref{eq10}, we get the following ranking function
\begin{equation}\label{eq15} 
  P(R=1 | d, q) \propto_R \sum_{\forall i:F_{i} = 1} \log  \frac{P(\mathit{tf}_i|E_{i}=1)}{\sum_{E_{i}\in\{0,1\}}P(\mathit{tf}_i|E_{i}) P(E_{i}) }
\end{equation}  
In order to estimate the probabilities in Eq. \eqref{eq15}, we assume that the collection of documents is a two component mixture for any given property. As $\xi$ is a set of 
term-description properties, we assume that the term frequency of the term associated with each property in $\xi$,
follows a Poisson distribution in a set of documents that are described by the property, $P(\mathit{tf}_i|E_{i}=1) \equiv e^{-\mu_{i}(1)} \mu_{i}(1)^{\mathit{tf}_i}$, and 
another Poisson distribution in the other set, $P(\mathit{tf}_i|E_{i}=0)\equiv e^{-\mu_{i}(0)} \mu_{i}(0)^{\mathit{tf}_i}$, where $\mu_{i}(1)$ and $\mu_{i}(0)$ are the two Poisson means. 
The mixing probability $P(E_i=1) \equiv p_i$ is an additional parameter. This is the classic 2-Poisson mixture model  \cite{harter75probabilistic,RobertsonRP80} with parameters $\mu_{i}(1), \mu_{i}(0), p_i$. 

For inference in the above mixture model, we can approach either in a maximum likelihood (ML) or in a Bayesian framework coupled with Markov Chain Monte Carlo (MCMC)  technique. 
For the experiments in the following section, we estimate the optimal parameter values of the mixture by using maximum likelihood 
estimation (using Expectation Maximization (EM) algorithm \cite{dempster77}) as well as Gibbs sampling for finite mixtures via MCMC~\cite{diebold1994estimation}. 
By substituting the estimated parameter values in Eq.\eqref{eq10}, we get the final ranking function
\begin{equation}\label{retrank}
 P(R=1 | d, q)  \propto_R  \\ \sum_{\forall i:F_{i} = 1} log \Big( \frac{e^{-\mu_{i}(1)} \mu_{i}(1)^{\mathit{tf}_i}} {p_{i} e^{- \mu_{i}(1)} \mu_{i}(1)^{\mathit{tf}_i} +  (1-p_{i}) e^{-\mu_{i}(0)} \mu_{i}(0)^{\mathit{tf}_i} }\Big)
\end{equation}
The ranking function in Eq.\eqref{retrank} looks similar to the ranking functions in \cite{RobertsonRP80, RobertsonW:1994} but is substantially different; the apparent similarity arises only from the use of the 
two Poisson distributional assumption.  
\subsection{Collaborative Filtering (CF)}
The unified probabilistic model in Section \ref{gmodel} can be directly used to rank and recommend a set of items for a given user once we define the properties that describe user, item and 
their relationships ($Y, Z$). So, in this section, we derive a ranking function specific to collaborative filtering (CF), when the only available information 
is the user-item rating matrix. 

Before deriving a relevance ranking model for CF, we describe the elements of the model in outline as follows:
(1) Each individual user is assumed to have preferences for certain kinds of items, similar to our example employer who seeks a candidate with certain characteristics 
for a job vacancy.  As we have initially no external indication of what `kinds' of items exist, this preference function is an unknown over the entire item space.  
That is, each item has a preference value for this user -- not as an individual item, but as a representative of `items like this'; (2) In an exactly dual form, 
each individual item is assumed to have appeal to different kinds of users.  Each user has an appeal value for this item -- not as in individual user, 
but as a representative of `users like this one'. (3) When an individual user sees an individual item, his/her reaction (rating) is assumed to be a stochastic function of the combination 
of user-item preference and item-user appeal.

In this version of the model, in the absence of any other properties, the `properties' of users are associated with individual items -- e.g. `this is an example of the kind of 
item that I like'.  The function of the matrix $Z$ is to map this back to users -- in other words to identify other users who like similar things.  Thus in this case the function of $Z$ identified 
at the end of section \ref{sec:hyp} becomes very clear.  Similarly, the properties of items are associated with individual users, and the matrix $Y$ performs the dual mapping.

\textbf{Relevance under the hypothesis:} Based on our hypothesis, a user $\textbf{u}$ and item $\textbf{i}$ pair is relevant if and only if:  the ``kinds of users'' to whom item $\textbf{i}$ appeals 
prefer the ``kinds of items'' preferred by $\textbf{u}$ and  the ``kinds of items'' preferred by user $\textbf{u}$ appeal to the ``kinds of users'' that the item $\textbf{i}$ appeals to.
The above definition of relevance is same as our general definition of relevance under the hypothesis. The only difference is the terminology 
replacing the need, item properties with ``kinds of items'' and ``kinds of users'' respectively. \footnote{i.e. the user is represented by the kinds of items he prefers and item is 
represented by the kinds of users it appeals to.}

Now, we derive a ranking function for CF using Eq.\eqref{eq3}.  In CF, the only available information about the users and items consists of 
the user-ids, the item-ids, and a set of ratings. Using this information, we define $I_p$ as the set of properties with one per item-id, i.e. each item is a different kind of 
item. A given user has a preference for the properties in $I_p$, where $I_p = \{i_{id_1}, i_{id_2}, \cdots, i_{id_N}\}$. In other-words, a user is described using $I_p$ properties. 
Similarly, we define $U_a$ as the set properties with one property per user-id, i.e. there are $M$ different kinds of users. And, a given item has an appeal factor to each user, 
where $U_a = \{u_{id_1}, u_{id_2}, \cdots, u_{id_M}\}$, i.e an item is described using $U_a$.\footnote{$I_p, U_a$ are same as $N$, $T$ in general model.}

Let $\mathit{Pf}$ be the preference matrix (i.e. $Y$) representing the relationship between the ``kinds of items'' (user properties) to ``kinds of users'' (item properties) and similarly, $A$ 
be the appeal matrix ($Z$). Let $\mathit{Pf}_{m}$ be the $u_{id_m}$ property preference vector over $I_p$  where $m \in \{1, \cdots, M\}$ and $u_{id_m} \in U_a$, 
i.e the $n^{th}$ entry in $\mathit{Pf}_{m}$ ($\mathit{Pf}_{mn}$) represents a binary value and equals to ``1'' if  $u_{id_{m}}$ prefers $i_{id_{n}}$, ``0'' otherwise. 
Similarly, $A_{n}$ be the $i_{id_n}$ property appeal vector over $U_a$, i.e the $m^{th}$ entry in $A_{n}$ ($A_{nm}$) is binary value and equals to ``1'' if $i_{id_{n}}$ appeal 
to users with property $u_{id_{m}}$. By substituting the above values in the Eq.~\eqref{eq5} we get, 
\begin{multline}\label{eq51}
P(R=1| u_{id}, i_{id}) = P(R=1)  \sum_{\alpha} \sum_{\beta} \sum_{\gamma} \sum_{\delta}  \prod_n \prod_m
 \frac{P(E_{n}, F_{m}, \mathit{Pf}_{mn}, A_{id_{nm}}| R=1)}{P(E_{n}) P(F_{m}) } \\ P(E_{n}|u) P(F_{m}| i)
\end{multline}
From the hypothesis, we know that if a user-item pair $(u_{m}, i_{n})$ is relevant then $E_{n} = 1$,  $F_m = 1$, $\mathit{Pf}_{mn} = 1$ and $A_{nm} = 1$. In other words, a user with a property 
$u_m$ prefers an item $i$ with property $i_{id_n}$, an $i$ appeals to a user with property $u_{id_m}$ and $u_{id_m}$ prefers the kind of item $i_{id_n}$,  kind of item $i_{id_n}$ appeal to the 
kind of user $u_{id_m}$. From the above assumption, if a user-item pair $u$, $i$ is relevant then $P(E_n=1, F_m=1,\mathit{Pf}_{mn}=1, A_{nm}=1| R=1) = 1$.%
\footnote{This assumption forces\\ $P(E_{n}, F_{m},\mathit{Pf}_{mn}, A_{nm}| R=1)= 0$ if any of the values $E_{n}, F_{m},\mathit{Pf}_{mn}, A_{nm}$ is zero for a relevant $u$, $i$ pair.} 
By substituting these values in Eq.\eqref{eq3}, we get, 
\begin{multline}\label{eq11}
P(R=1| u, i) =  P(R=1)  \prod_{<n_r, m_r>} \frac{P(E_{n_r}=1|u) P(F_{m_r}=1| i)}{P(E_{n_r}=1) P(F_{m_r}=1)}  
\sum_{\alpha^{'}} \sum_{\beta^{'}} \sum_{\gamma^{'}} \sum_{\delta^{'}} \\ \prod_{<n_{nr}, m_{nr}>} P( E_{n_{nr}}, F_{m_{nr}}, 
Pf_{{m_{nr}}{n_{nr}}}, A_{{n_{nr}}{m_{nr}}}| R=1) \\ \frac{P(E_{n_{nr}}|u) P(F_{m_{nr}}| i)}{P(E_{n_{nr}}) P(F_{m_{nr}}) } 
\end{multline}
where $n_r, m_r$ such that $<u_{m_r}, i_{n_r}> \in UI_{rel}$ where $UI_{rel}$ is the set of relevant $<u, i>$ pairs. Similarly, $n_{nr}, m_{nr}$ such that 
$<u_{m_{nr}}, i_{n_{nr}}> \notin UI_{rel}$. Now, we make an assumption that we have only a set of relevant user item pairs and then by approximating and removing the 
 constant $P(R=1)$ from the Eq. \eqref{eq11}, we get,  
\begin{equation}\label{eq13}
P(R=1| u, i) \propto_R \prod_{<n_r, m_r>} \frac{P(E_{n_r}=1|u) P(F_{m_r}=1| i)}{P(E_{n_r}=1) P(F_{m_r}=1)}
\end{equation}
To estimate preference and appeal in CF model, we make an assumption that an observed rating of a user-item pair, $<u, i>$, has a stochastic element associated with the 
item's appeal to the kind of user $u$ belongs to and the user's preference for the kind of item that $i$ belongs to.  

In order to estimate the preference distribution of an individual user over kinds of item,  we further assume that this user's observed ratings are the result \emph{only} of this user's preferences.
Similarly, to estimate the appeal distribution of an individual item over kinds of user, 
we assume that the observed ratings on this item are the result \emph{only} of this item's appeal. These two assumptions are clearly oversimplifications but more sophisticated models 
can be pursued in future work.  

Using the assumptions, we compute the probability that the kind of item $i_{id}$ appeals to the user $u$ as  $P(E_{i_{id}}=1| u) \equiv P(E_{i_{id}}=1| r)$ 
where $r$ denotes the observed user $u$ rating on item $i$. Similarly, we compute the probability that the kind of user  $u_{id}$ prefers the item $i$ as 
$ P(F_{u_{id}}=1| i) \equiv P(F_{u_{id}}=1| r)$. We make another assumption that, ratings, $r$,  given by a kind of user $u_{id}$ to a set of items follows 
one distribution in the kind of items s/he prefers and another distribution in non-preferred kind of items. Similarly, the rating $r$ received by a kind of  item  $i_{id}$ follows one 
distribution in the kind of user that the item appeals to and another distribution in the ratings received from the kind of users it does not appeal to. Therefore, we can draw a 
probabilistic inference about the preference of a user from his associated ratings over a set of items. By applying Bayes' rule to $P(E_{i_{id}}=1| r)$, we get 
\begin{equation} \label{eq20}
P(E_{i_{id}}=1| r) = \frac{P(r|E_{i_{id}}=1) P(E_{i_{id}}=1)}{\sum_{E_{i_{id}} \in\{0,1\}} P(r|E_{i_{id}}) P(E_{i_{id}})}.
\end{equation}
Similarly, we compute $P(F_{u_{id}}=1| r)$. 
By substituting the above values in Eq.\eqref{eq13},  we get the ranking function,
\begin{multline} \label{rank}
 P(R=1| u,i) \propto_R  \prod_{<n_r, m_r>} \frac{P(r_{ u {i_{n_r}}}|E_{i_{n_r}}=1)}{\sum_{E_{i_{n_r}}\in\{0,1\}}P(r_{ui_{n_r}}|E_{i_{n_r}}) P(E_{i_{n_r}})}
 \\ \frac{P(r_{ i u_{m_r}}| F_{u_{m_r}}=1)}{ \sum_{ F_{u_{m_r}} \in\{0,1\} } P(r_{i u_{m_r}}|F_{u_{m_r}}) P( F_{u_{m_r}})  }.
\end{multline} 
where $\propto_R$ is rank equivalence and $r_{u {i_{n_r}}}$ is the observed user $u$ rating on $i_{n_{r}}^{th}$ kind of item, similarly, $r_{i u_{m_r}}$ is 
the observed $i$'s received rating from the kind of user $u_{m_r}$.

To compute the probabilities in Eq.~\eqref{rank}, we can use a version of the 2-Poisson mixture used for ad-hoc retrieval.
We assume that the item's received ratings from the kinds of users to which it appeals follow a Poisson distribution, and a different 
Poisson distribution among users to whom it does not appeal. We make a similar assumption about the ratings of a user. 
Thus we have $\mu_{i_{n_r}}(1)$ and $\mu_{i_{n_r}}(0)$, two Poisson means of ratings received by the kind of item  $i_{n_r}$, and a mixing 
probability $P(E_{i_{id_{n_r}}}) \equiv p_{i_{n_r}}$, and a two Poisson mixture for each item's ratings, with parameters  $\mu_{u_{m_r}}(1)$ 
and $\mu_{u_{m_r}}(0)$ and the mixing probability $p_{u_{m_r}}$.\footnote{Although ratings on a scale [1-5] are not the same as term frequencies, the fact 
that they are small integers makes the 2-Poisson assumption work passably well.}

By substituting the parameter values in Eq.\eqref{rank} and applying logarithm, we get the final ranking function as 
\begin{equation}\label{cfrank}
   P(R=1| u,i) \propto_R  \sum_{<n_r, m_r>}  \log  \frac{A}{ p_{i_{n_r}} A +(1- p_{i_{n_r}}) B} 
\\+ log \frac{C}{ p_{u_{m_r}} C +  (1-p_{u_{m_r}}) D} 
\end{equation}
where\\
\ensuremath{
\small
\begin{array}{lll} 
A =  e^{-\mu_{i_{n_r}}(1)} \mu_{i_{n_r}}(1)^{r_{u{i_{n_r}}}}, B = e^{-\mu_{i_{n_r}}(0)} \mu_{i_{n_r}}(0)^{r_{u {i_{n_r}}}}, 
C = e^{-\mu_{u_{m_r}}(1)} \mu_{u_{m_r}}(1)^{r_{iu_{m_r}}}, \\ 	D =e^{-\mu_{u_{m_r}}(0)} \mu_{u_{m_r}}(0)^{r_{i u_{m_r}}}.
\end{array}
}\\

Eq.~\eqref{cfrank} is the final collaborative filtering ranking function, which making use of related user-item pairs to perform the calculation. It is important to note that 
unlike the Matrix Factorization methods and dimension reduction methods, such as SVD \cite{SVD:Koren} and topic models \cite{Hofmann:2004} to name just a few, we do not need to set 
any specific number of hidden dimensions in which both the users and items will be represented.  In other words, it does not involve a lower dimensional representation of features. 
Also, there is no need to compute explicitly the similarities between the users or items, which is the basis of the user-based approaches \cite{Shardanand95} and the item-based 
approaches \cite{Sarwar:2001:ICF:371920.372071}.

\begin{figure}[t]
\centering
  \subfigure[Used ratings]{\label{fig:cf:intuition:ratings}\includegraphics[width=0.39\textwidth]{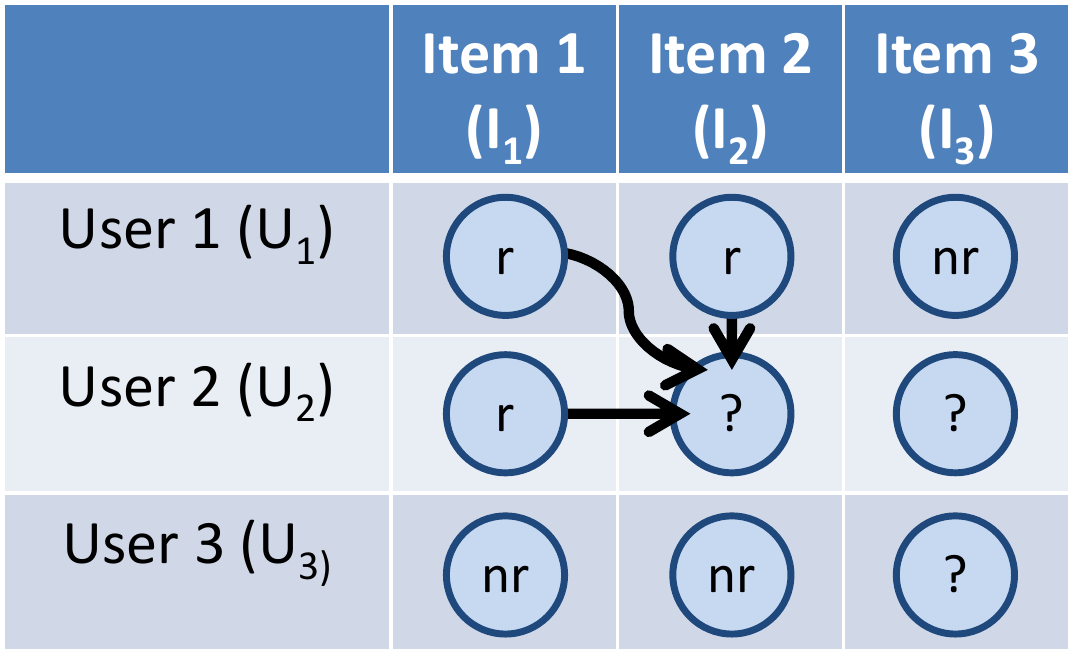}}
  \subfigure[Used user-item pairs]{\label{fig:cf:intuition:pairs}\includegraphics[width=0.39\textwidth]{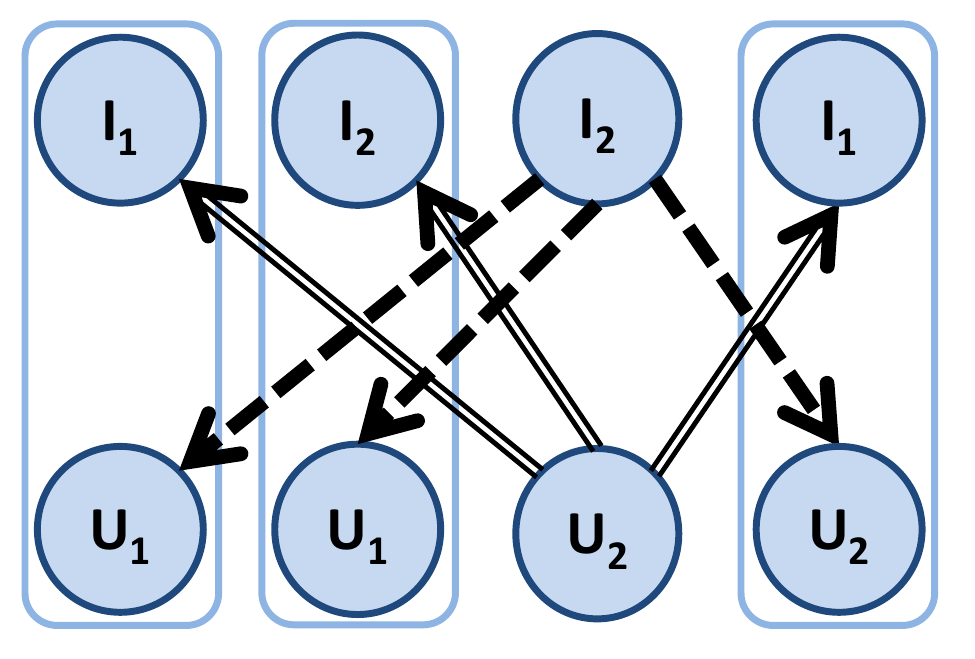}} 
  \caption{Relevance Propagation in the Unified Model. The solid lines indicate the \emph{preference} of $U_{2}$ to the `kind of items' and the dotted lines 
  indicate the \emph{appeal} of $I_{2}$ to the `kind of users'.}
\label{tab:gt}
\end{figure}

Instead, our method explores implicit similarity by computing the $u$'s \emph{preference} to a `kind of item' and $i$'s \emph{appeal} to a `kind of user' in a relevant 
user-item pair as shown in Fig. \ref{tab:gt} (b). By combining the \emph{preference} and \emph{appeal} of user-item pair, the relevance information of a relevant user-item 
pair will be propagated to the relevance between the $u, i$ pair; this is illustrated in Fig. \ref{tab:gt} (a). That is, if $u$ likes a different item which also appeals to 
another user who likes item $i$, then these known relevant pairs will affect the probability of $u$, $i$ being relevant. This is different compared to a unified collaborative 
filtering model presented in \cite{wang:tois}, where an unknown rating is estimated by explicitly similarity measures from three sources: the user's own ratings for different 
items (item-based), other user's ratings for the same item (user-based), and, ratings from different but similar users for other but similar items.

 \section{Experiments}
In this section we present our results on Ad-Hoc retrieval only. 
 \subsection{Test Collections}
 The objective of our experiments is to see how well the resulting rankings functions in Eq.\eqref{retrank} and Eq.\eqref{cfrank} perform in the text retrieval and 
 collaborative filtering (CF) applications respectively. For the ad-hoc retrieval evaluation, we used five different TREC document collections, representing small 
 to medium sizes: 1) FBIS on disc 5, 2) Financial Times (FT) on disk 4, 3) LosAngeles Times (LA) on disk 5, 4) TREC-7 and TREC-8 ad hoc retrieval document collection, 
 Disk 4 \& 5 minus Congressional Record, and 5) WT10G collection. The topic sets used are: 1) topics 301-350 , 2) topics 401-450, 3) topics 501-550 and 
 4) topics 301-350 and 601-700 minus 672. We use the document collection followed by the TREC number as a label for the test collection, e.g. FBIS-8 represents the test 
 collection with FBIS document collection and TREC-8 topics (i.e. 401-450). Similarly, labels, \emph{Robust}, \emph{TREC-10} represents the TREC, 4\&5 document collection 
 with Robust topics and WT10G collection with topics 501-550 respectively. For each of these collection queries are formed from the title field only. 

\begin{table*}[t]\footnotesize
 \centering
 \small\addtolength{\tabcolsep}{-2pt}
 \caption{Comparison of the Unified Model with other baseline models. $t$-$test$ with 95\% confidence is used and the statistically significant results (with respect to the second best models) are marked with $\dagger$.}
 {\small\begin{tabular}{|c|c|c|c|c|c|c|c|c|}
 \cline{1-7}
 & & \multicolumn{5}{|c|}{Model Name \& Performance} \\ \cline{1-7}
 Collection & Metric & BM25 & LM - JM & Dirichlet-LM & UM (EM) & UM (Bayesian) \\ \cline{1-7}
 \multicolumn{1}{|c|}{\multirow{1}{*}{FT-8}} &
 \multicolumn{1}{|c|}{MAP} & 0.323 & 0.317 &  0.325 & 0.347 & \textbf{0.347}$\dagger$ \\ \cline{2-7}
 \multicolumn{1}{|c|}{}                        &
 \multicolumn{1}{|c|}{MRR} & 0.649 & 0.590 & 0.664 & 0.711 &\textbf{0.724}$\dagger$   \\ \cline{1-7}
 \multicolumn{1}{|c|}{\multirow{1}{*}{FBIS-8}} &
 \multicolumn{1}{|c|}{MAP} & 0.326 & 0.306 &  0.325 & 0.315 & \textbf{0.334}$\dagger$ \\ \cline{2-7}
 \multicolumn{1}{|c|}{}                        &
 \multicolumn{1}{|c|}{MRR} & 0.598 & 0.496 & 0.598 & 0.560 & \textbf{0.614}$\dagger$   \\ \cline{1-7}
 \multicolumn{1}{|c|}{\multirow{1}{*}{LA-8}} &
 \multicolumn{1}{|c|}{MAP} & 0.254 & 0.232 &  0.256 & 0.260 &\textbf{0.276}$\dagger$ \\ \cline{2-7}
 \multicolumn{1}{|c|}{}                        &
 \multicolumn{1}{|c|}{MRR} & 0.565 & 0.402 & 0.545 & 0.583 &\textbf{0.594}$\dagger$   \\ \cline{1-7}
 \multicolumn{1}{|c|}{\multirow{1}{*}{TREC-8}} &
 \multicolumn{1}{|c|}{MAP} & 0.251 & 0.239 &  0.256 & 0.257 &\textbf{0.260} \\ \cline{2-7}
 \multicolumn{1}{|c|}{}                        &
 \multicolumn{1}{|c|}{MRR} & 0.644 & 0.476 & 0.638 & 0.654 &\textbf{0.670}$\dagger$   \\ \cline{1-7}
 \multicolumn{1}{|c|}{\multirow{1}{*}{TREC-7}} &
 \multicolumn{1}{|c|}{MAP} & 0.193 & 0.180 &  0.192 & 0.191 &\textbf{0.195}\\ \cline{2-7}
 \multicolumn{1}{|c|}{}                        &
 \multicolumn{1}{|c|}{MRR} & 0.652 & 0.551 & 0.650 & 0.630 &\textbf{0.667}   \\ \cline{1-7}
 \multicolumn{1}{|c|}{\multirow{1}{*}{Robust}} &
 \multicolumn{1}{|c|}{MAP} & 0.242 & 0.185 &  0.245 & 0.245 &\textbf{0.248} \\ \cline{2-7}
 \multicolumn{1}{|c|}{}                        &
 \multicolumn{1}{|c|}{MRR} & 0.650 & 0.564 & \textbf{0.668} & 0.620 &0.638   \\ \cline{1-7}
 \multicolumn{1}{|c|}{\multirow{1}{*}{TREC-10}} &
 \multicolumn{1}{|c|}{MAP} & 0.193 & 0.148 &  0.193 & 0.190 & \textbf{0.195}\\ \cline{2-7}
 \multicolumn{1}{|c|}{}                        &
 \multicolumn{1}{|c|}{MRR} & 0.596 & 0.451 & 0.588  & 0.60 & \textbf{0.611}  \\ \cline{1-7}
 \end{tabular}}
 \label{tab:performance}
 \vspace{10pt}
 \end{table*}

We also initialized the mixture parameters by using the collection statistics as follows: For the EM algorithm, we initialized $p$ as the percentage of the 
documents where the term occurs.  Thus the initial rank function is equivalent to the IDF weighting (see the discussion in Section 3). We used a minuscule value to 
initialize $\mu(0)$ by assuming that the average term frequency of a term associated with the term-description elite property in a document approaches zero if it is non-elite to 
the document. Similarly, $\mu(1)$ was initialized with the average number of times the term appeared in document collection with its term frequency in a document more than one.  
For the Gibbs sampling, we chose the prior parameters values in the similar fashion.
 
\textbf{Performance:} After learning the parameters from each document collection, we employed our ranking function in Eq.~\eqref{retrank} on each test collection and 
computed the performance metric scores. Table \ref{tab:performance} summarizes the results of the unified ranking function (UM), using EM \& Bayesian estimation, along with 
the results of the baselines. The labels LM-JM corresponds to the language modeling ranking method \cite{Ponte98alanguage} with Jelinek-Mercer smoothing, whereas Dirichlet-LM 
corresponds to the Language Model with Dirichlet prior. From Table ~\ref{tab:performance}, we can see that our ranking function outperforms other models in most cases (some of 
them are significant). Because the rank function does not use any information other than the term statistics in document collection, we believe the improvement was due to the 
term-based parameters estimation, similar to the per-term smoothing in the Poisson based query-generation language models \cite{Mei:2007}.  Moreover, the performance of our model 
on title queries is comparable to the improved reported results in \cite{Mei:2007}.  In summary, the ad-hoc retrieval experiments show that the unified retrieval theory has great 
potential in text retrieval. A simple ranking function derived from our unified theory demonstrates that it can handle the retrieval situation without relevance feedback.

\section{Conclusion}
We have presented a new unified theory for information retrieval. We considered retrieval as a matching problem between two sets of properties, 
one from information needs and one from information items. To estimate the probability of relevance between them, we argued that the retrieval system not only needs 
to identify which item properties are `sought' by each need, but also to identify which need properties are `sought' by each item.  
We validated the proposed theory by formulating and developing practical relevance ranking functions for both ad-hoc text retrieval and collaborative filtering. 
We evaluated ad-hoc retrival ranking function performance on publicly available test collections (TREC collections for ad-hoc retrieval task).  
Besides the theoretical contribution, our experiments demonstrated its wide applicability.

There are fruitful avenues for future investigations into the proposed unified retrieval framework. For instance, we intend to extend and test the current text 
retrieval rank function and apply it to web search where relevance information is available (in the form of click-through data). It is of great interest to study the 
theory in other IR applications such as content filtering, multimedia retrieval, people matching and search, opinion retrieval \cite{Huang:opinion}, and advertising.

\bibliographystyle{acm}
\bibliography{sigproc}
\end{document}